\documentclass[11pt,twoside]{article}

\usepackage{asp2006}
\usepackage[pdftex]{graphicx}
\usepackage{lscape}
\usepackage [compact]{titlesec}
\usepackage{paralist}

\titlespacing{\section}{6pt}{0ex}{0ex}

\begin{document}
\title{Updating the Historical Sunspot Record}
\author{Leif Svalgaard}
\affil{HEPL, Via Ortega, Stanford University, Stanford, CA 94305-4085, USA}

\begin{abstract}
We review the evidence for the argument that Rudolf Wolf's calibration of
the Sunspot Number is likely to be correct and that Max Waldmeier introduced
an upwards jump in the sunspot number in 1945. The combined effect of these
adjustments suggests that there has been no secular change in the sunspot
number since coming out of the Maunder Minimum $\sim$1715.
\end{abstract} 

\section{The Sunspot Record(s!)}
The Sunspot Record goes back 400 years and is the basis for many reconstructions
of solar parameters (e.g. TSI), but, how good is it? And can we agree on which
one (Wolf Number, International Number, `Boulder' Number, Group Number, ...)? Are
the old values good? Are the new ones? And what is a `good' or `correct' Sunspot
Number anyway?

Johann Rudolf \citet{wo1859} defined his Relative Sunspot Number, taking into account
both individual spots and their appearance in distinct groups (what we today call
`active regions'), as \(R_W=10\,Groups+Spots\). Wolf started his own observations
in 1849 and assembled observations from earlier observers back to 1749 and beyond
(Figure 1).

\begin{figure}[!ht]
\includegraphics [width=\textwidth]{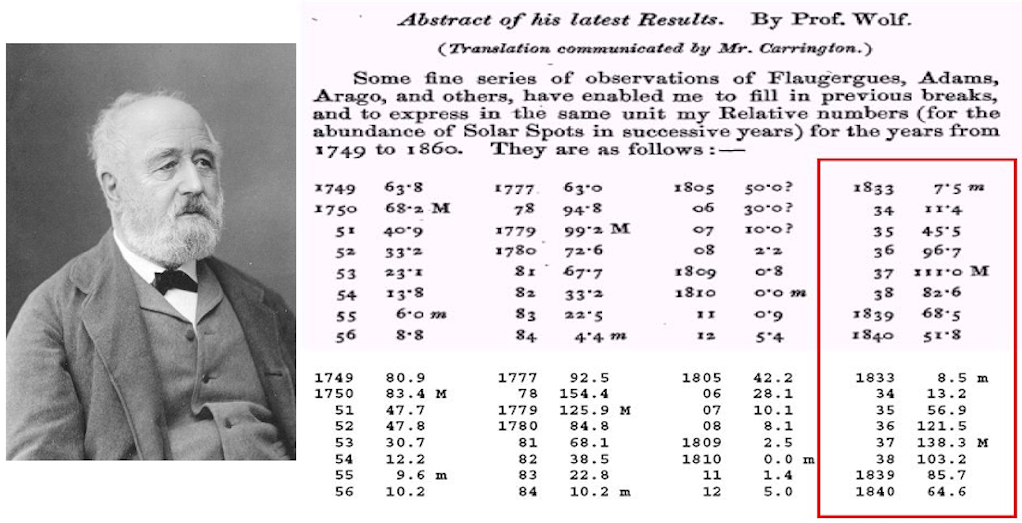}
\caption{Rudolf Wolf and excerpts from his 1861 list of published Relative Numbers
compared with his latest list (now the official list from SIDC in Brussels).}
\end{figure}

As is clear, the earlier values were subsequently adjusted (upwards) as Wolf were
struggling with the difficulty of bringing different observers onto the same `scale',
compensating for telescope size, counting method, acuity, seeing, and personal bias.

Wolf published several versions of his celebrated Relative Sunspot Numbers based on
data gathered from many observers from both before and during Wolf's own lifetime
(Figure 2). How to `harmonize' data from different observers?

\begin{figure}[!ht]
\includegraphics [width=\textwidth]{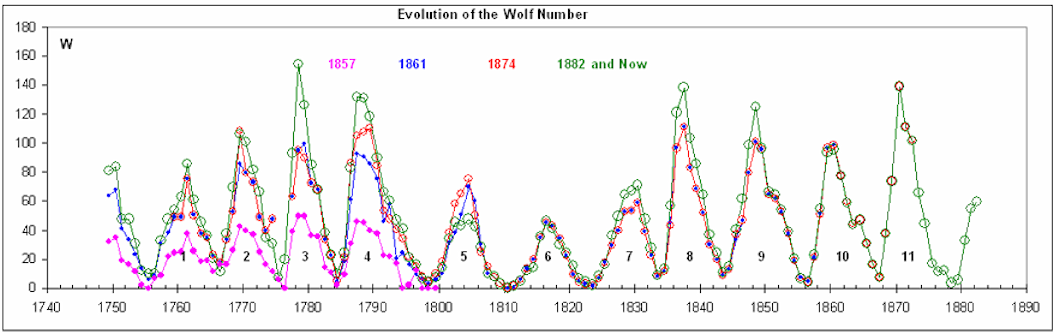}
\caption{Evolution of the Wolf Number from his first 1857 list to the final version,
with color coded symbols and curves for each list} 
\end{figure}

\section{Wolf's Elegant Solution}
A current system in the ionospheric E-layer is created and maintained by solar UV
radiation (Figure 3). The current has a magnetic field of its own which is readily
observed on the ground even with 18th Century technology. This variation
was, in fact, discovered in 1722 by George \citet{gr1724} as a regular variation of
typically 10 arc minutes during each day of the angle (called the Declination today)
a compass needle makes with true North. The amplitude (\textit{rD}) of this variation 
is an excellent proxy for solar UV.

\begin{figure}[!ht]
\includegraphics [width=\textwidth]{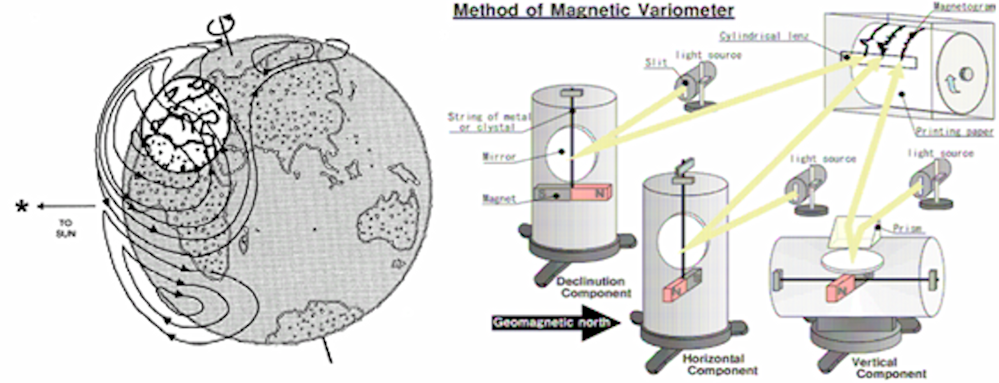}
\caption{Ionospheric current system (left) fixed with respect to the Sun. Stations rotate
into and out of the magnetic field of this system, recording it (right).}
\end{figure}

\citet{wo1859} discovered that this amplitude has a strong linear relationship with his
newly defined Relative Number, \textit{$R_W$}: \(rD=a+b\,R_W\) and used the 
relationship to calibrate the sunspot number on a yearly basis (Figure 4). 

\vspace{0.2in}\noindent
Wolf  made two overall major calibration changes based on \textit{rD}:
\noindent
\begin{compactitem}
\item (1861) Sunspot numbers before $\sim$1798 were doubled  
\item (1875) All values before 1849 (including the ones that were doubled) 
were increased by 25\%
\end{compactitem}

\begin{figure}[!ht]
\includegraphics [width=\textwidth]{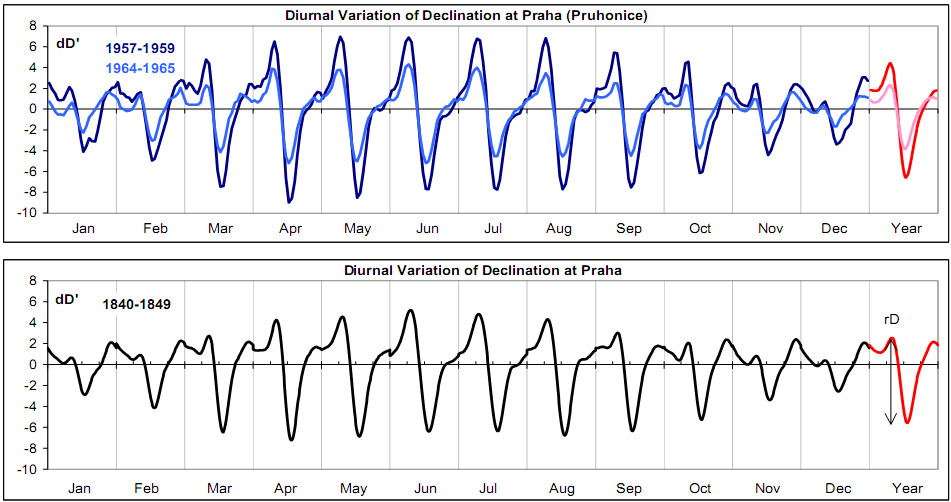}
\caption{Diurnal variation of Declination at Prague per month. Top: modern data for
low sunspot number (light blue) and for high sunspot number (dark blue). Bottom: for
the 1840s. The amplitude changes with the solar zenith angle. Wolf used the yearly
average (red) to calibrate the yearly \textit{R$_W$}.}
\end{figure}

By comparing sunspot numbers (SSN) reported by other observers with his own, Wolf
introduced a scale factor to compensate for the differences: \(SSN=k_W(10\,Groups+Spots)\).

\vspace{0.1in}
\section{The Group Sunspot Number}
\citet{hs1998} proposed basing the Sunspot number solely on the number of
groups reported by the observers: \(GSN = 12\,Groups\). The calibration constant was
used to make the value of the GSN comparable to the modern Sunspot Number. However,
the number of sunspot groups is also observer dependent (Figure 5), by up to a factor
of two or more, so an observer-dependent adjustment factor is also needed:
\(GSN=12\,k_G\,Groups \).

\begin{figure}[!ht]
\includegraphics [width=\textwidth]{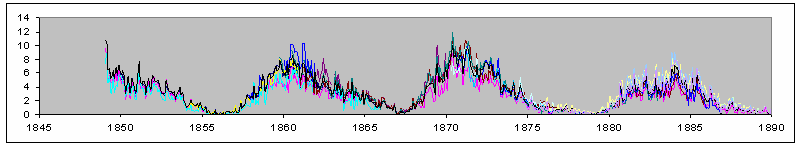}
\caption{Group counts by several observers: Schwabe, Wolf, Carrington, Shea, Peters,
Sp\"orer, Weber, Schmidt, Secchi, Bernaerts, Wolfer, Aguilar, Ricco, and RGO, shown with
different color coding.}
\end{figure}

So, the conceptually cleaner GSN also needs adjustment and `bridges' from one observer
to the next, and the next, etc, lacking the `absolute' calibration afforded by another
physical observable. This may lead to a possibly spurious secular change (Figure 6).

\begin{figure}[!ht]
\includegraphics [width=\textwidth]{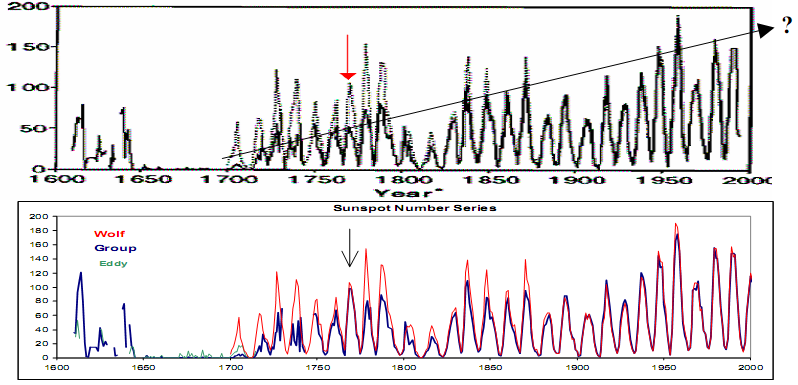}
\caption{Two versions of the GSN (top: GSN full line, Wolf SSN dotted line). The latest
GSN is compared to the Wolf (and International) SSNs in the bottom panel. The GSN is
largely based on the RGO (Greenwich) dataset from 1875 onwards.}
\end{figure}

\section{Sunspot Number Relationship with Diurnal Range}
Extensive datasets exist (\citet{sc1909}) from the `Magnetic Crusade' in the 1840s and 
for times after the First Polar Year 1882 (Figure 7). 

\begin{figure}[!ht]
\includegraphics [width=\textwidth]{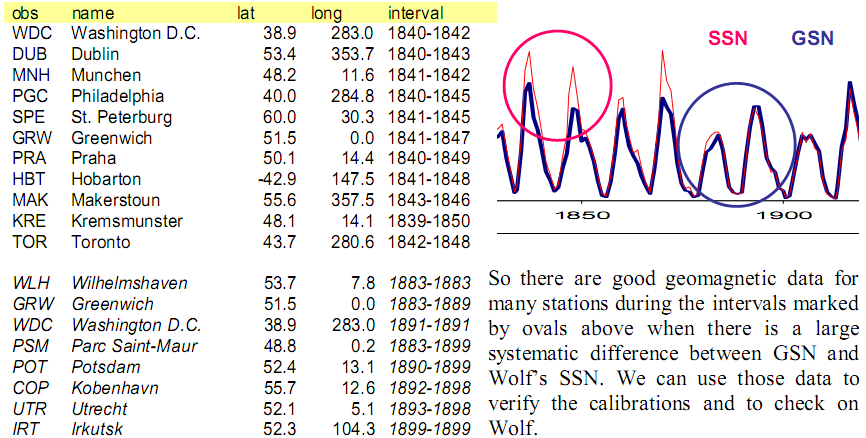}
\caption{List of stations for which we know amplitudes of\textit{rD} for the intervals indicated}
\end{figure}

As the current flows along meridians on the morning and evening sides, the magnetic
deflection is along latitude circles and the magnitude in force units (nT) is largely
constant from station to station over a wide latitude range (20\deg-60\deg). From the 
observed values of \textit{rD} and of the Horizontal component, \textit{H}, the range in 
force units,\textit{ rY}, is readily calculated as shown. Wolf didn't know that the important 
parameter is \textit{rY} and not \textit{rD}, so he had to contend with regression coefficients 
that varied from station to station and with time. While not a serious problem once you know 
why, this variation nevertheless weakened other researchers' confidence in Wolf's procedure.

As shown in Figure 8, after $\sim$1882 the GSN (pink) and the SSN (blue) have the same relation
with the range of geomagnetic variation and cluster neatly along a common regression line,
also found for SSN for stations regardless of time interval.

\begin{figure}[!ht]
\includegraphics [width=\textwidth]{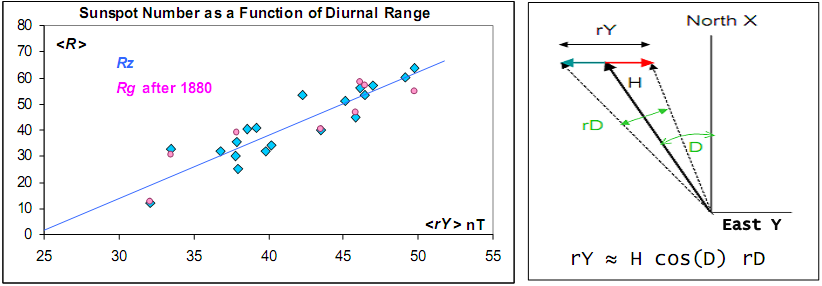}
\caption{Average sunspot numbers (GSNs in pink and Wolf SSNs in blue) for the intervals
in the table above vs.\ the average range of the diurnal variation of the East Component
of the geomagnetic field, \textit{rY}.}
\end{figure}

If we add the GSN averages (red diamonds) for stations before 1850, we find that
they all fall well below the regression line for stations after 1880 (Figure 9).

\begin{figure}[!ht]
\center{\includegraphics [width=256pt]{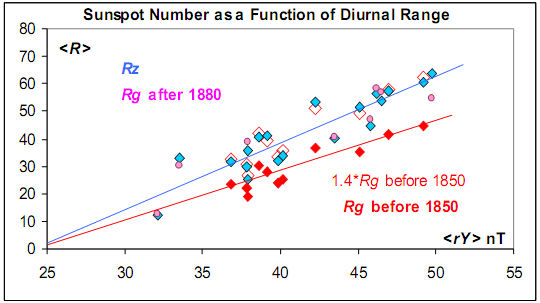}}
\caption{Same as Figure 8, but with the average GSNs for intervals before 1850s added
(filled red diamonds). Scaling these values up by a factor of 1.4 (open red diamonds)
brings them into fair agreement with the common regression line for the Wolf SSNs and
for the GSNs after 1880.}
\end{figure}

If the diurnal range, \textit{rY}, is a satisfactory measure of the kind of solar activity we
believe the sunspot should be a proxy for, then the above analysis would suggest that the 
Group Sunspot Number should be increased by a factor of 1.4 sometime before $\sim$1880,
removing most of the discrepancy between the two sunspot number series.

\vspace{0.1in}
\section{Ratio of Wolf SSN and Group Sunspot Number}
It is instructive to plot the ratio of the Wolf number and the Group number (omitting
years where either is close to zero). Figure 10 shows that with the above adjustment by
a factor of 1.4 before $\sim$1880, the ratio seems to be near unity, with an expected large 
noise component early on. A discontinuity in 1945 when Max Waldmeier took over production 
of the Z\"urich Sunspot Number is apparent. Alfred Wolfer became Wolf's assistant around 
1880 and began to influence the `Wolf' number from then on.

\begin{figure}[!ht]
\includegraphics [width=\textwidth]{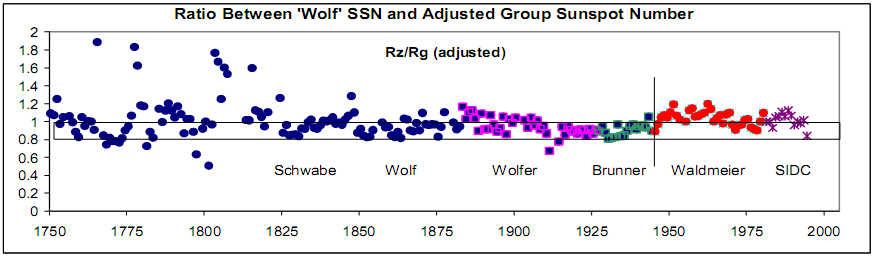}
\caption{Ratio of yearly values of the Wolf (Z\"urich) Sunspot Number, \textit{R$_Z$}, and 
the Group Sunspot Number, \textit{R$_G$}. Different observers are indicated by appropriate 
color and symbol coding.}
\end{figure}

Wolf did not count pores and the smallest spots. His assistant (and successor) Alfred
Wolfer disagreed and argued that all visible pores and spots, no matter how small, should be counted,
and of course won the argument by staying longer on the right side of the grass. He used the
correction factor, \textit{k$_W$} = 0.6, to bring his counts into conformance with Wolf's. There is 
some confusion about \textit{k$_W$} being a `personal' factor or not. As Wolfer used it, \textit{k$_W$} 
compensates for a difference in what is counted, rather than a difference in telescope, seeing, etc. 
Since the number of groups is rarely influenced by a change in counting the smallest spots, the
defining equation should perhaps better have been \(SSN=10\,Groups+\kappa\,Spots \), but this
is now probably too late to change.

\vspace{0.1in}
\section{Geomagnetic Ranges; the Waldmeier Discontinuity, I}
\citet{fr2005} comments on the change of observers in Z\"urich in 1945 and writes: "The new
observer-team was thus relatively inexperienced" and "Waldmeier himself feared that his scale
factor could vary". We now know that his fear was not unfounded. Waldmeier's counts are 22\%
higher than Wolfer and Brunner's, for the same amplitude of the Diurnal Geomagnetic Variation
(\mbox{Figure} 11). This is close to the size of the discontinuity deduced from Figure 10. As SIDC
took pains to maintain continuity with Waldmeier, the jump carries over to modern SSN values.

\begin{figure}[!ht]
\includegraphics [width=\textwidth]{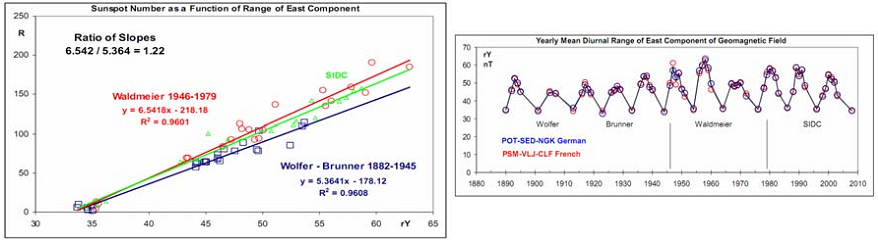}
\caption{Select years near solar minimum and near solar maximum, then plot the yearly Z\"urich
SSN as a function of the range of the East component for two station chains, one German and 
one French.}
\end{figure}

\section{The RGO Sunspot Area Series; the Waldmeier Discontinuity, II}
There is a strong correlation (with zero offset) between the Sunspot area (\textit{SA}) and
\(R_Z=(1/r) SA^{0.775}\). The ratio \(r=SA^{0.775}/R_Z\) is observer dependent. Histograms
of the ratio values indicate that Waldmeier's \textit{R$_Z$} values are a factor of 3.39/2.88 = 1.18
too high (Figure 12), or 18\%.

\begin{figure}[!ht]
\includegraphics [width=\textwidth]{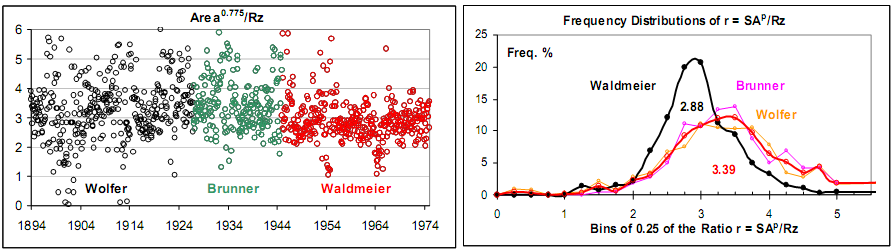}
\caption{Monthly values of the ratio between the sunspot areas (to power \textit{p} = 0.775) for
different observers as indicated.}
\end{figure}

\section{CaK Spectroheliograms; the Waldmeier Discontinuity, III}
From $\sim$40,000 CaK spectroheliograms from the 60-foot tower at Mount Wilson between 1915 and
1985 a daily index of the fractional area of the visible solar disk occupied by plages and active 
network has been constructed \citet{be2008}. Monthly averages of this index are strongly correlated with the sunspot number. The relationship is not linear, but can be represented by the equation: 
\(R_Z = $[$(CaK-0.002167)*8999)$]$^{1.29}\) using data from 1915-1945, i.e. the pre-Waldmeier 
era. The SSN reported by Waldmeier is $\sim$20\% higher than that calculated from CaK using the above pre-Waldmeier relation, as can be seen in Figure 13.
\citet{fo1998} reports a similar result.

\begin{figure}[!ht]
\includegraphics [width=\textwidth]{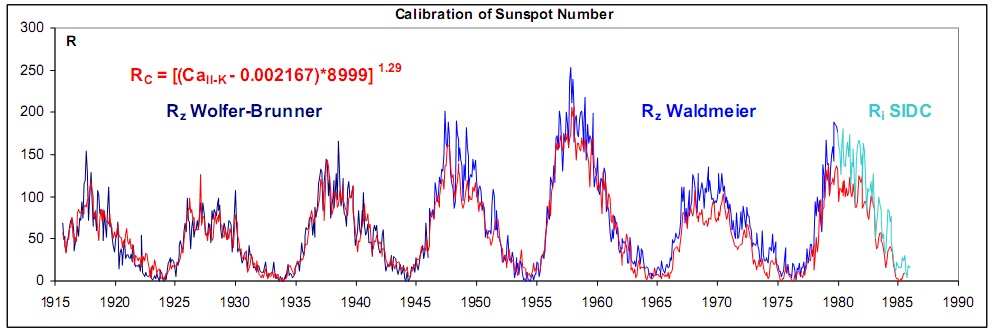}
\caption{Comparison of Zurich SSN (blue colors) and a synthetic SSN (red) calculated from a
Ca II K-line index using a pre-Waldmeier relationship.}
\end{figure}

\vspace{0.1in}
\section{Ionospheric Critical Frequency; the Waldmeier Discontinuity, IV}
The value of the Ionospheric Critical Frequency \textit{foF2} depends strongly on solar activity
(\citet{ph1948}) and is perhaps the parameter with the clearest response to solar activity.
The slope of the correlation changed 24\% between solar cycles 17 and 18 when Waldmeier took
over, corresponding to a 24\% higher SSN after 1945 than that which would be expected from
the \textit{foF2} relation. Based on these several lines of independent evidence there seems little
doubt that Waldmeier introduced a spurious increase of the Z\"urich sunspot number of that
magnitude.

\vspace{0.1in}
\section{Geomagnetic Range is an Excellent Proxy for F10.7 Radio Flux}
Wolf's linear relationship (on which the calibration hangs) is completely vindicated by modern
data using the F10.7 cm flux as a measure for general solar activity. The coefficient of
determination \textit{R$^2$} is in excess of 0.98 for yearly averages of the flux (itself a proxy 
for solar UV) and the amplitude of the ensuing geomagnetic diurnal variation (Figure 14). 
This establishes that Wolf's procedure and calibration are physically sound and precise enough 
to be applicable.

\begin{figure}[!ht]
\includegraphics [width=\textwidth]{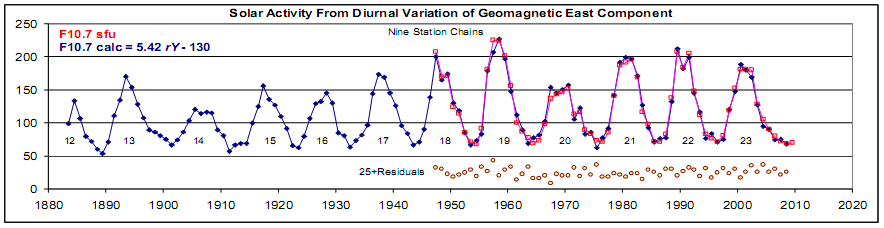}
\caption{The yearly average range, \textit{rY}, derived from nine long-running chains of geomagnetic
observatories (blue) compared with the average F10.7 cm solar flux (red).}
\end{figure}

\section{Conclusion}
Modern data shows that the diurnal range of the geomagnetic variation is an extremely good proxy
for the solar microwave flux. To the extent that we take the flux to be a measure of general
solar activity of which the sunspot number was meant to be an indicator, we argue that Wolf's
calibration makes his sunspot series essentially an equivalent F10.7 series. Accepting the
soundness of Wolf's procedure and correcting for the Waldmeier discontinuity (+20\%) lead to a
picture of solar activity with but little difference between activity levels in the 18$^{th}$, 19$^{th}$,
and 20$^{th}$ centuries (Figure 15).

\begin{figure}[!ht]
\includegraphics [width=\textwidth]{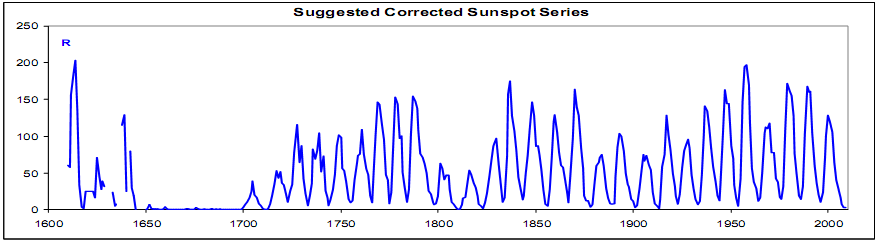}
\caption{Suggested equivalent sunspot numbers calibrated by the geomagnetic record.}
\end{figure}

\end{document}